\documentclass[reprint,
superscriptaddress,
nofootinbib,
amsmath,amssymb,
twocolumn
]{article}
\usepackage{authblk}

\usepackage{CJK}
\usepackage{graphicx}
\usepackage{dcolumn}
\usepackage{bm}
\usepackage{hyperref}
\usepackage{url} 
\usepackage{color}					
\hypersetup{colorlinks=true,allcolors=blue}

\usepackage[mathlines]{lineno}
\usepackage{makeidx} 
\usepackage{natbib}
\usepackage{xfrac}
\usepackage{multirow}
\usepackage[export]{adjustbox} 
\usepackage{comment}
\usepackage{float}
\usepackage{etoolbox}
\usepackage{tabularx}

\usepackage{tikz}
\newrobustcmd*{\mycircle}[1]{\tikz{\filldraw[draw=black,fill=#1] (0,0) circle [radius=0.07cm];}}
\newrobustcmd*{\mytriangle}[1]{\tikz{\filldraw[draw=#1,fill=#1] (0,0) -- (0.14cm,0) -- (0.07cm,0.14cm);}}

\usepackage{dcolumn}
\usepackage{gensymb}
\usepackage{color}
\usepackage[caption = false]{subfig}

\usepackage{mathptmx}

\begin{document}

\title{Isomeric fission yield ratios for odd-mass Cd and In isotopes using the Phase-Imaging Ion-Cyclotron-Resonance technique}

\author[1]{V. Rakopoulos \footnote{e-mail: \href{mailto:Vasileios.Rakopoulos@physics.uu.se}{Vasileios.Rakopoulos@physics.uu.se}}}
\author[1]{M. Lantz \footnote{Corresponding author: \href{mailto:Mattias.Lantz@physics.uu.se}{Mattias.Lantz@physics.uu.se}}}

\author[1]{S. Pomp}%

\author[1]{A. Solders}%

\author[1]{A. Al-Adili}%

\author[2]{L.~Canete}%

 \author[2]{T.~Eronen}%

 \author[2]{A.~Jokinen}%

 \author[2]{A.~Kankainen}%

\author[1]{A. Mattera}%
 
 \author[2]{I.~D.~Moore}%
 
 \author[2]{D.~A.~Nesterenko}%

 \author[2]{M.~Reponen}%

 \author[2]{S. Rinta-Antila}%

 \author[2]{A. de Roubin}%

 \author[2]{M. Vil\'{e}n}%

\author[1]{M.~\"{O}sterlund}%

 \author[2]{H.~Penttil\"{a}}%
\affil[1]{Department of Physics and Astronomy, Uppsala University, Uppsala 75120, Sweden}
\affil[2]{Department of Physics, FI-40014 University of Jyv\"{a}skyl\"{a}, Finland}

\date{}

\twocolumn[
  \begin{@twocolumnfalse}
    \maketitle
      \begin{abstract}
Isomeric yield ratios for the odd-$A$ isotopes of $^{119-127}$Cd and $^{119-127}$In from 25-MeV proton-induced fission on natural uranium have been measured at the JYFLTRAP double Penning trap, by employing the Phase-Imaging Ion-Cyclotron-Resonance technique.  With the significantly improved mass resolution of this novel method isomeric states separated by 140 keV from the ground state, and with half-lives of the order of 500 ms, could be resolved. This opens the door for obtaining new information on low-lying isomers, of importance for nuclear structure, fission and astrophysics. In the present work the experimental isomeric yield ratios are used for the estimation of the root-mean-square angular momentum ($J_\mathrm{rms}$) of the primary fragments. The results show a dependency on the number of unpaired protons and neutrons, where the odd-$Z$ In isotopes carry larger angular momenta. The deduced values of $J_\mathrm{rms}$ display a linear relationship when compared with the electric quadrupole moments of the fission products.
 \end{abstract}
\end{@twocolumnfalse}
]

Fission fragments carry a considerable amount of angular momentum~\cite{Vandenbosch1973,PhysRevC.5.2041}, but it is still a puzzling question how it is generated~\cite{Andreyev2017}.~Different theories compete on the interpretation of this issue, and among others involve thermal excitation~\cite{PhysRevC.60.034613,Shneidman2003,Gonnenwein2007} and/or quantum-mechanical uncertainty of angular-momentum-bearing modes~\cite{MORETTO1989453,PhysRevC.75.064313}, Coulomb excitation after scission~\cite{PhysRev.133.B714} and strong coupling between the elongation and other collective degrees of freedom~\cite{PhysRevLett.116.122504}.~At higher excitation energies, extra unpaired nucleons that exist at the saddle point of the highly excited nucleus contribute additionally to the fragments' spin. Furthermore, a partial retainment of the initial angular momentum by the fragments can be assumed as the result of excitations corresponding to collective modes where the two nascent fragments move relative to one another~\cite{Cuninghame80}.

Currently, there is no direct way of measuring the angular momentum of the primary fission fragments. Thus, isomeric yield ratios can provide an important tool, as being one of the fission observables from which the angular momentum can be inferred~\cite{PhysRev.120.1305,PhysRev.120.1313}.~The highly excited fragments created in fission de-excite by emitting neutrons and $\gamma$-rays, before they are eventually trapped at the isomer or populate the ground state.~During this process, the average angular momentum of the primary fission fragments plays an important role, as it controls the number of emitted neutrons and photons.~Usually low initial angular momenta produce many neutrons and few $\gamma$-rays, and vice versa~\cite{PhysRevC.88.044603}.~Based on the statistical equilibrium among various collective modes~\cite{NIX19651} and according to the pre-scission bending mode oscillation model~\cite{RASMUSSEN1969465}, the average angular momentum of the primary fragment can be related to the temperature, neck radius and deformation of the fragments at the scission~point~\cite{Datta1986}.

Isomers are created due to a combination of factors, such as shape, spin and spin projection, that inhibit their decay, resulting in excited states with lifetimes longer than the usual states~\cite{Walker:1999aa,Dracoulis13,Dracoulis16}.~The unique properties of isomers are of interest in a range of potential applications, such as energy storage and $\gamma$-ray lasers~\cite{Walker:1999aa}, studies of coupled atomic-nuclear effects~\cite{Becker06,MATINYAN1998199,GOBET201180,PhysRevLett.99.172502,Chiara2018}, and in medical diagnostics~\cite{Walker05} and treatment~\cite{Kassis2011}.~Moreover, knowledge on isomers and their de-excitation can illuminate many nuclear phenomena related to nuclear structure, as presented in Refs.\cite{Dracoulis13,KONDEV201550} and references therein, and astrophysics studies.~For example, in the astrophysical rapid neutron capture process, fission cycles material back to lower-mass regions~\cite{RevModPhys.29.547,ARNOULD200797}.~Population of isomeric states can affect the final abundances due to their different $\beta$-decay half-lives, as well as $\beta$-delayed neutron and neutron-capture probabilities.~Thus, nuclides in the region near the $``magic"$ tin, such as the Cd and In isotopes studied in the present work, are of particular interest. 

In addition, the knowledge of the direct yield of metastable states in fission can be necessary for nuclear energy applications.~The $\beta$-delayed neutron emission probabilities may be notably different from that of the ground state.~Combined with the long half-lives of the isomeric states, these two properties have an impact on criticality and decay heat in reactors.~While the principles about the latter are well established, exact knowledge on the $\beta$-decay feeding probability of specific contributors to the heating of the reactors still lacks sufficiently accurate information~\cite{PhysRevLett.105.202501}.~As some metastable states are among them, precise determination of their yields can be of importance for reactor safety, economy, and the efficient use of available resources~\cite{Yoshida2007}.~Isomer ratios can also be of importance to experiments related to antineutrino spectra generated by nuclear reactors, which are calculated based on fission yield data and isomeric ratios~\cite{PhysRevC.91.011301,PhysRevLett.116.132502,PhysRevLett.115.102503}.~Specifically, the products of $^{238}$U contribute disproportionately to these spectra and may  be responsible for the observed shoulder in these spectra~\cite{PhysRevLett.118.222501,PhysRevD.92.033015}. Lastly, the evolution of isomeric yield ratios as a function of mass $A$ can be of interest for improving the models that predict the isomeric yield ratio through de-excitation calculations that depend on, among others, the angular momentum and parity of the primary fission fragments.

Based on the isomeric yield ratios the angular momentum of the primary fission fragments can be deduced~\cite{PhysRev.138.B353,PhysRevC.16.160,PhysRevC.30.195,Fujiwara1982,IMANISHI1976141,PhysRevC.38.1787}, combined  with the statistical model analysis,  first introduced by Vandenbosch $et~al.$~\cite{PhysRev.120.1305,PhysRev.120.1313}. It can also be estimated by measuring other fission observables, such as the angular distribution of prompt $\gamma$ rays from the fission fragments~\cite{PhysRev.133.B714,Strutinsky1960,HAMILTON1997273,Urban1997,PhysRevC.60.064611}, the energy and multiplicity of prompt $\gamma$ rays~\cite{PhysRevC.6.1023,JOHANSSON1964378,JOHANSSON1965147,NIFENECKER1972285}, and the intensities of the cascade transitions to the ground state of the rotational bands~\cite{PhysRevC.5.2041}. However, the method based on isomeric yield ratio measurements can be used for all fission products, irrespective of their $Z$ and $A$ number, while the other methods exhibit certain limitations~\cite{rakopoulos18}. Furthermore,  isomer production ratios have been employed in investigations of collective rotational degrees of freedom~\cite{Cuninghame80,PhysRevC.71.014304,Datta1986,PhysRevC.28.1206}.

Isomeric yield ratios have been experimentally determined by means of $\gamma$ spectroscopy, either by applying radiochemical separation~\cite{IMANISHI1976141,NAIK1995273} or by physical means using an ordinary isotope separator (selection of A)~\cite{Bail:2011aa}. By producing and separating the fission products with the Ion Guide Isotope Separator On-Line (IGISOL) technique~\cite{AYSTO2001477}, any obstacles such as insufficient mass resolving power in the separator, difficulties to study refractory elements, and inadequate knowledge of decay schemes, that usually encounter the aforementioned experimental techniques, can be overcome.~The half-lives of the nuclides remain a constraint for all techniques.

In this work, the isomeric yield ratios of the odd-mass isotopes of In and Cd in the mass range $A = $~119 - 127 have been experimentally determined.~The odd-$N$ isotopes of Cd ($Z$=48), typically have ground states of spin 1/2$^{+}$ or 3/2$^{+}$ and isomeric states of spin 11/2$^{-}$, due to the shell configuration of a neutron hole in 2d$_{3/2}$ or 3s$_{1/2}$ and 1h${_{11/2}}$, respectively.~The even-$N$ isotopes of In ($Z$=49), have ground states of spin 9/2$^{+}$ and isomeric states of spin 1/2$^{-}$, corresponding to shell configurations of a proton hole in the shells 1g$_{9/2}$ and 2p$_{1/2}$ (see Table~\ref{tab:level}). The excitation energies of the isomers studied in this work span over the range \mbox{147-409 keV}, with the exception of the second isomeric state in $^{127}$In, which is at 1870 keV and could be observed simultaneously with the first isomeric state and the ground state. 

The fission fragments were produced at the IGISOL facility~\cite{AYSTO2001477}, by a 25-MeV proton beam impinging on a 15~mg/cm$^{2}$ thick $^\mathrm{nat}$U target.~The fission products are thermalised in a helium buffer gas, extracted by a sextupole ion guide~\cite{Karvonen:2008aa} and electrostatically accelerated to 30q keV (where q is the charge state of the ions, usually q=1).~A dipole magnet is used for mass separation of the isobaric chain of interest based on the mass-to-charge ratio (m/q).~The continuous mass-separated beam is directed into the radiofrequency cooler and buncher (RFQ)~\cite{PhysRevLett.88.094801,Nieminen:2001aa}, where the ions are collected, cooled and subsequently injected as a short bunch into the JYFLTRAP double Penning trap mass spectrometer~\cite{Eronen2012}.

\begin{figure}[h]
\centering
\subfloat{\includegraphics[scale=0.2]{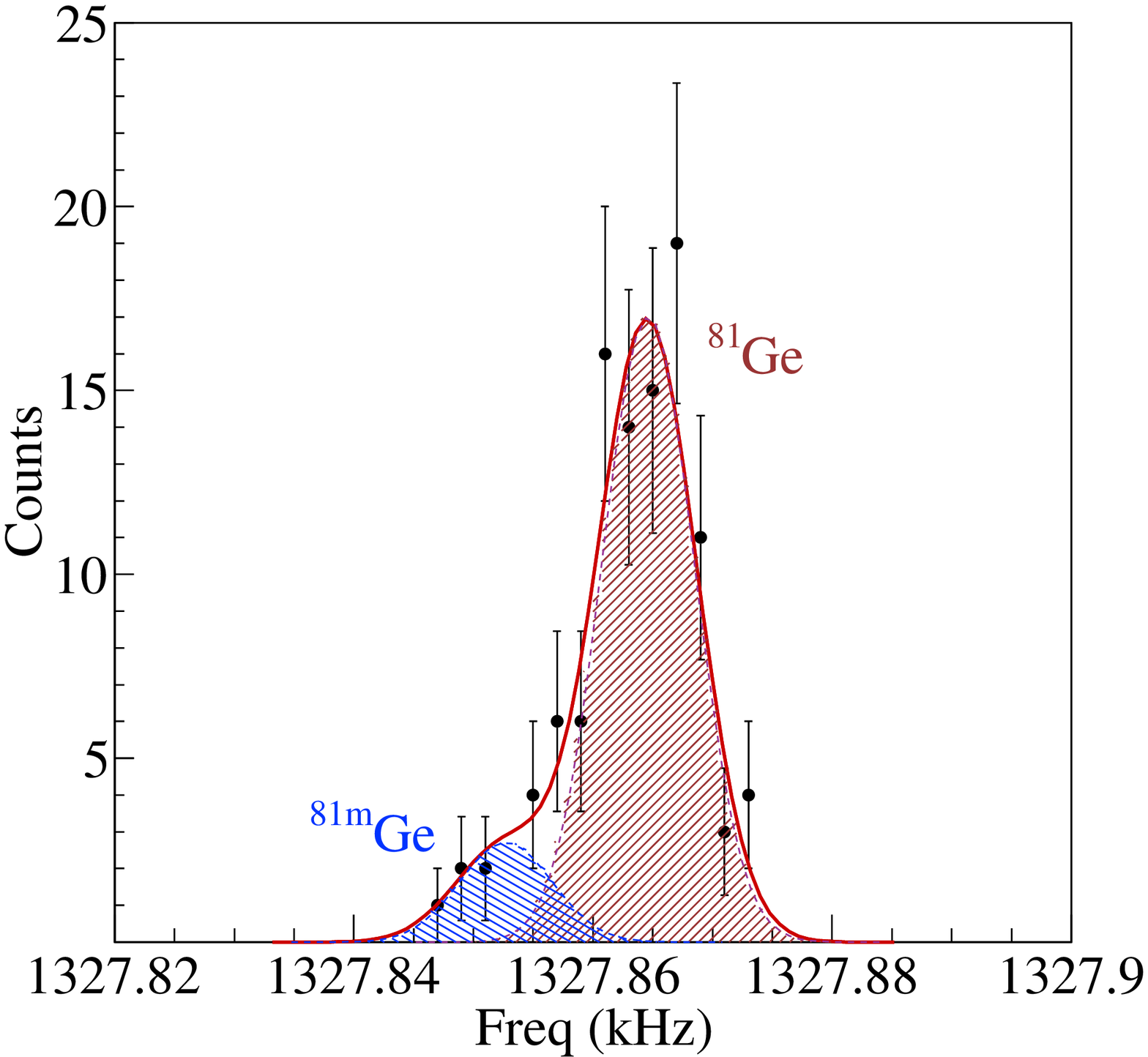}}
\subfloat{\includegraphics[scale=0.2]{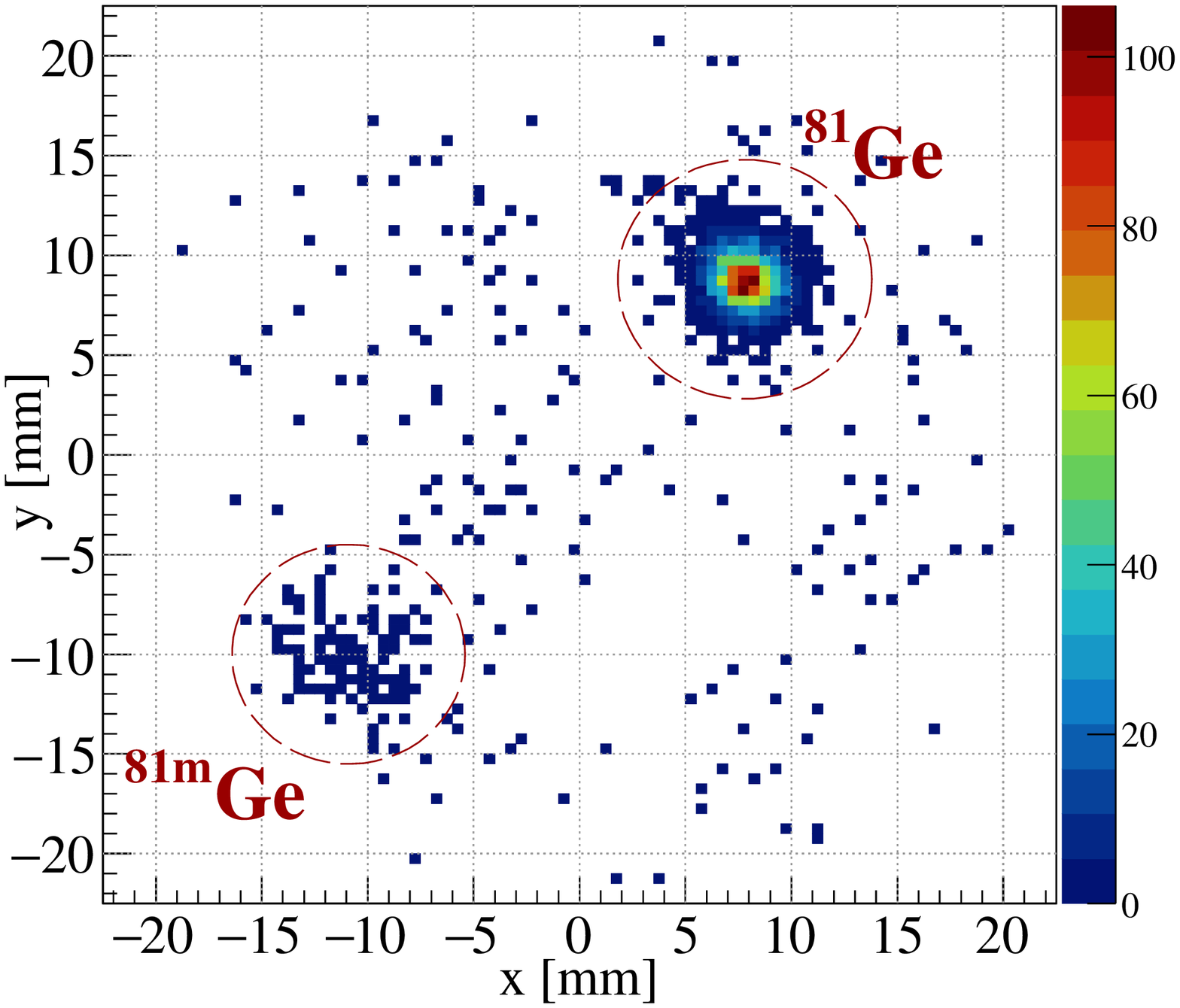}}
\caption{The case of $^{81}$Ge, observed with the sideband cooling technique (left) and with the PI-ICR technique (right). The dashed circles are used to guide the eye.}
\label{81Ge}
\end{figure}

The isomeric yield ratios were determined at JYFLTRAP by employing the recently implemented Phase-Imaging Ion-Cyclotron-Resonance (PI-ICR) technique~\cite{Nesterenko18}, first introduced by Eliseev $et~al.$~\cite{PhysRevLett.110.082501}, for mass measurements of short-lived nuclides.~The mass resolving power of this novel technique is significantly improved compared to the one previously used at JYFLTRAP~\cite{rakopoulos18}, where the isomeric yield ratios were determined by employing the sideband cooling technique~\cite{SAVARD1991247}.~Thus, with the PI-ICR method, states with energy separation as low as 100 keV can be resolved. The superior resolving power is highlighted in Fig.~\ref{81Ge} for the case of $^{81}$Ge, where in the lefthand plot the isomeric yield ratio is determined with the sideband cooling technique~\cite{rakopoulos18}, while in the righthand plot the PI-ICR technique is employed.~It is evident that while in the former case the mass resolving power limit is reached, the new technique allows a complete separation of the states. Improved mass resolution, shorter measurements cycles, direct ion counting and independency of knowledge of decay schemes  enable the study of a large number of isomers of relevance for the different applications mentioned above. 

Ion samples from the states of interest are initially prepared in the so-called purification trap by applying the sideband cooling technique~\cite{SAVARD1991247} in order to remove the isobars, and only keep the states of interest.~Afterwards, the ions are injected into the precision trap. The PI-ICR technique starts with the application of a dipolar pulse at the mass-dependent modified cyclotron frequency $\nu_+$ of the radial motion so that the ions are moved to a radius of about 10 mm.~The excitation pulse is chosen to be short ($\sim$1 ms) so that ions within  $\pm$500 Hz are excited, ensuring equally strong excitation for the close lying isomers.~After an excitation-free evolution time $t_{acc}$, the ions have accumulated a total cyclotron phase of $\phi+2\pi n=2\pi\nu_+t_{acc}$, where $n$ is the number of full revolutions that the ions perform during the accumulation time.~Due to the mass-dependency of the frequency $\nu_+$, separation of the states of interest can be achieved, as the phase advance of each state will be different at the end of the accumulation time. In the last step, a quadrupolar pulse at the sum frequency ($\nu_+ + \nu_-$) is applied in order to convert the cyclotron motion to magnetron motion~\cite{Nesterenko18}.~This pulse is fast ($\sim$2 ms), corresponding to a width of $ \nu_c$ of about $\pm$250 Hz. In this measurement, a maximum separation of 180$\degree$ between the states was chosen, dictating the accumulation time $t_{acc}$.~For the case of $^{127}$In, where three states could be identified, a suitable separation was chosen so that overlap of the states could be avoided.

\begin{figure*}[t]
\centering
\includegraphics[scale=0.8]{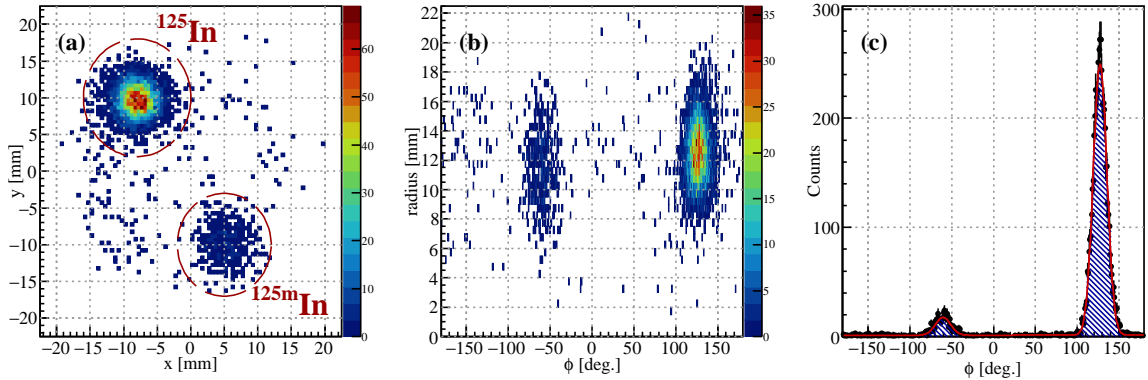}
\caption{A typical histogram showing the position of the two states of $^{125}$In, separated by 180$\degree$ (a). The next two panels illustrate the analysis procedure followed in this work. In (b), the position of the ions are converted to polar coordinates. In (c), the yield of each state is determined from gaussian fitting on the $\phi$ projection. The dashed circles are used to guide the eye.}
\label{analysis_step}
\end{figure*}

The ions are extracted from the Penning trap, and the image of their phase is registered in a position-sensitive microchannel plate ion detector with a delay line anode~\cite{MCP} (see Fig.~\ref{analysis_step}$\color{blue}\mathrm{a}$).~For the data analysis, the position of the ions is converted to polar coordinates (Fig.~\ref{analysis_step}$\color{blue}\mathrm{b}$), where the analysis is performed on the $\phi$ projection (Fig.~\ref{analysis_step}$\color{blue}\mathrm{c}$).~The peaks are fitted with Gaussian distributions with a background, and the yields are determined from the area of the fitted peaks.~A mirrored measurement for each isomeric pair was performed, with the position of the states swapped, in order to cancel out any inhomogeneities in the detector efficiency. Since the isomeric yield ratios of the two mirror measurements agree within statistical uncertainties, the reported isomeric yield ratio is the weighted mean of these two measurements.~For the peak identification, an additional measurement was performed after each set of measurements with a chosen separation between the peaks of 90$\degree$.~Since the direction (evolution) of the cyclotron motion, as projected onto the MCP, is counterclockwise, the peaks can be identified.

For the most short-lived nuclides corrections to the observed yields had to be applied to account for radioactive decay losses that occur from the creation of the ions until their detection.~During the measurement, the ions are assumed to accumulate in the RFQ at a constant rate. Depending on the half-life and the yield, ions were accumulated in the cooler for 22-512 ms before being injected to the JYFLTRAP Penning trap.~In the Penning trap, only radioactive decay losses occur.~In the present experiment, the total time the ions spend in the Penning traps varies from 420 to 922 ms.~Consequently, these two time periods, which are the most time-consuming stages of the experiment, need to be taken into consideration for the applied corrections, as explained more in detail in Ref.~\cite{Penttila2010}. These corrections also contribute to the overall uncertainty of the observed yield mainly because of the uncertainties of the half-lives of the states. 

In Table~\ref{tab:level} the experimentally determined isomeric yield ratios, defined as the high spin state yield over the total yield, and the angular momentum of the primary fragments, as derived by using the nuclear reaction code {\footnotesize TALYS}~\cite{Koning:2007}, are reported.~A detailed description of the method employed to estimate the angular momentum is presented in Refs.~\cite{rakopoulos18,Rakopoulos_thesis}. The uncertainties of the isomeric yield ratios include the statistical uncertainty, as well as the uncertainty due to applied corrections. The uncertainties of $J\mathrm{_{rms}}$ are derived from the experimental ones.

\begin{table*}[t]
\caption{Nuclear properties of the nuclides studied in this work. Spin-parity (I$^{\pi}$), half-life (T$_{1/2}$) and excitation energy (E$_{x}$) are provided for each state. The metastable state is the higher spin state for all the isotopes of Cd and $^{129}$Sb, while the ground state is the higher spin for all the isotopes of In and $^{81}$Ge.~$^{127}$In is the only case where two isomeric states can be observed. In the spin-parity columns the parenthesis indicate uncertainty in the given values.~The values denoted by ``{\scriptsize \#}'' are estimated from trends in neighbouring nuclides with the same $Z$ and $N$ parities.~All data are retrieved from N{\scriptsize UBASE}2016~\cite{Audi16}.~The experimentally determined isomeric yield ratios (IYR), defined as the high-spin state yield over the total yield of the isotope, for the 25-MeV proton-induced fission on $\mathrm{^{nat}U}$ are given in the penultimate column.~The deduced values  for the average root-mean-square angular momentum ($J_\mathrm{rms}$) are reported in the last column.~The half-lives of $^{127}$Cd were kindly provided by C.~Lorenz through private communication~\cite{Lorenz2018}. Note that since for $^{127}$In two isomeric states could be observed, the isomeric yield ratio corresponds to the yield of the two highest spin states over the total yield.}
\setlength\tabcolsep{5pt}

\begin{tabularx}{\linewidth}{llllllll}
  
    \hline\hline

 \\ [-0.3cm]
 &  \multicolumn{2}{c}{Ground state} & \multicolumn{3}{c}{Isomeric state} \\ [0.09cm] 
\cline{2-3} \cline{4-6} \\ [-0.3cm]

Nuclide & ~~I$^{\pi}$ & ~~T$_{1/2}$& ~~~I$^{\pi}$ & ~~T$_{1/2}$ &~E$_{x}$ (keV) & ~~IYR & ~~$J^{av}_\mathrm{rms}$\\ [0.07cm] 
\hline
\\ [-0.3cm]

$^{81}$Ge & 9/2$^+${\scriptsize \#} & 8 (2) s &  (1/2$^{+}$) & 8 (2) s & 679.14 (4) & 0.975 (7) \\ [0.07cm]  

$^{119}$Cd & 1/2$^+$ &  2.69 (2) m  &11/2$^-$ & 2.20 (2) m & 146.54 (11) & 0.871 (15) & 12.3 (5)\\ [0.07cm] 
$^{121}$Cd & 3/2$^+$ & 13.5 (3) s & 11/2$^-$ & 8.3 (8) s & 214.86 (15) & 0.867 (4) & 14.7 (1)\\ [0.07cm] 
$^{123}$Cd & 3/2$^+$ & 2.10 (2) s & 11/2$^-$ & 1.82 (3) s & 143 (4) & 0.876 (7) & 15.7 (2)\\ [0.07cm] 
$^{125}$Cd & 3/2$^+$ & 680 (40) ms &  11/2$^-$ & 480 (30) ms & 186 (4) & 0.902 (8)\\ [0.07cm] 
$^{127}$Cd & 3/2$^+$ & 360 (40) ms &  11/2$^-$ & 450 (120) ms & 276 (15) & 0.872 (38)\\ [0.07cm] 

$^{119}$In & 9/2$^+$ & 2.4 (1) m & 1/2$^-$ & 18.0 (3) m & 311.37 (3) & 0.978 (15) & 26.2 (4)\\ [0.07cm] 
$^{121}$In & 9/2$^+$ & 23.1 (6) s & 1/2$^-$ & 3.88 (10) m & 313.68 (7) & 0.971 (11) & 25.1 (5)\\ [0.07cm] 
$^{123}$In & (9/2)$^+$ & 6.17 (5) s & (1/2)$^-$ & 47.4 (4) s & 327.21 (4) & 0.958 (2) & 21.2 (2) \\ [0.07cm]
$^{125}$In & 9/2$^+$ & 2.36 (4) s &   (1/2)$^{(-)}$ & 12.2 (2) s & 360.12 (9) & 0.950 (3) & 15.9 (3)\\ [0.07cm]  
\multirow{2}{*}{$^{127}$In} & (9/2$^+$) & 1.09 (1) s & 1/2$^-${\scriptsize \#} & 3.67 (4) s & 408.9 (3) & \multirow{2}{*}{0.921 (2)} & \multirow{2}{*}{9.5 (2)}\\ [0.07cm]  
& & & (21/2$^-$) & 1.04 (10) s & 1870 (60)\\ [0.07cm] 

$^{129}$Sb & 7/2$^+$ & 4.366 (26) h & (19/2$^{-}$) & 17.7 (1) m & 1851.31 (6) & 0.441 (32) &  \\ [0.03cm]  

    \hline\hline
\end{tabularx}
\label{tab:level}
\end{table*}

In order to assure the consistency of the novel technique and verify the results for $^{81}$Ge and $^{129}$Sb, which have previously been measured at the same facility with the sideband cooling technique~\cite{rakopoulos18}, the measurements of these nuclides were repeated.  A good agreement between the two techniques is observed.  The isomeric yield ratios for $^{81}$Ge and $^{129}$Sb, as taken from Ref.~\cite{rakopoulos18}, are 0.97 (1) and 0.47~(4),~respectively.

For the case of $^{127}$In, where three states could be quantified, the ratio is estimated as the sum the yield of the two highest-spin states over the total yield of the isotope. The individual fractional yields are 62.4\% for the ground state, 7.9\% for the first isomer and 29.7\% for the second isomer. The isomeric yield ratios of the Cd isotopes are almost constant, with the exception of $^{125}$Cd where a noticeable increase can be seen. On the other hand, the yield ratios of the In isotopes show a monotonic decrease with respect to the mass number. Although there is a significant deviation between the isomeric yield ratios of the In and Cd at lower mass number, a tendency of decreasing difference can be observed as the mass number approaches the doubly magic region $A$~=~132.

\begin{figure}
\centering
\includegraphics[scale=0.4]{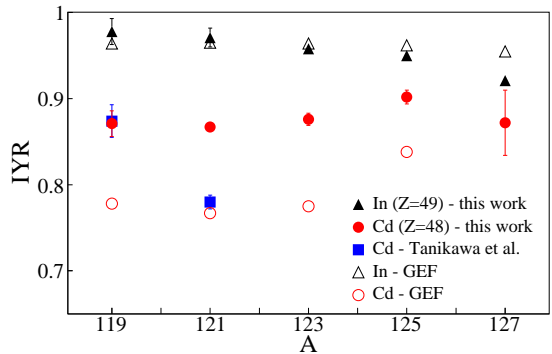}
\caption{The experimentally determined isomeric yield ratios for the isotopes of In and Cd as measured in this work by employing the PI-ICR technique.~The data are compared with data available in the literature~\cite{Tanikawa:aa}, and with results from calculations performed with the {\footnotesize GEF} model~\cite{Schmidt2014,SCHMIDT2016107}.~The error bars in the experimental results are smaller than the data points, whenever they are not visible in the figure.}
\label{Jrms}
\end{figure}

In Fig.~\ref{Jrms}, the experimental results are compared with calculations performed with the {\footnotesize GEF} model for ten million events, version~2017/1.2~\cite{Schmidt2014,SCHMIDT2016107} and experimental data obtained at the Tohoku IGISOL facility in Japan, by means of $\gamma$~spectroscopy~\cite{Tanikawa:aa}. An excellent agreement between the results of this work and the {\footnotesize GEF} model can be noticed for the In data, except for the case of $^{127}$In. Note that for this case {\footnotesize GEF} provides only the yield for the ground state and the first isomeric state, while in our case the fractional yield also includes the second isomer.~For the Cd isotopes, although the values from {\footnotesize GEF} are consistently lower, the trend is the same. For $^{127}$Cd, {\footnotesize GEF} does not give any result.~The comparison of the results of this work to the ones reported by~Tanikawa $et~al.$ shows an excellent agreement for the case of $^{119}$Cd, while for $^{121}$Cd our result deviates by about 9 standard deviations. 

The angular momentum of the primary fission fragments was derived by employing the nuclear reaction code {\footnotesize TALYS}~\cite{Koning:2007}.~In these calculations, by varying the angular momentum distribution of the primary fragments, an agreement was sought between the isomeric yield ratio as estimated by TALYS with the experimental results~\cite{rakopoulos18}.~The formation of a specific isomeric pair can result from contributions of various primary fragments, depending on the de-excitation path and the number of emitted neutrons.~In the present work, by taking into account the mass-dependent post-scission neutron multiplicity~\cite{Rubchenya:2001aa}, the contribution from primary fragments after emitting from one up to three neutrons are considered.~Thereafter, in order to estimate the average root-mean-square angular momentum, the individual results are weighted by the mass and charge distribution and the neutron emission probability of the primary fragments, as taken from the {\footnotesize GEF} model.

The {\footnotesize TALYS} code could match the experimental results for the cases of $^{119,121,125,127}$In and $^{119, 121,123}$Cd.~For $^{125,127}$Cd, the inability of the code to reproduce the experimental result is not surprising considering how poorly the level schemes of these isotopes are known.~For $^{123}$In, {\footnotesize TALYS} could meet the experimental result only when some of the experimentally known levels at high excitation energies were replaced by a level density model. Specifically, the discrete levels that were included in the calculations are those that constitute the~\textit{``complete''} scheme, according to the  RIPL-3 database~\cite{CAPOTE20093107}.~According to RIPL-3, this scheme contains 20 out of the 33 experimentally known levels.~This is an indication that some levels or $\gamma$-ray transitions between level 20 and 33 might be missing, which apparently affect the relative population of the isomeric~state.

The origins of $J_\mathrm{rms}$ of fission fragments can be regarded as a combination of contributions from collective degrees of freedom and single particle excitations. Other effects such as Coulomb excitation can also be relevant.~In the present work, the derived angular momenta of the In isotopes show a monotonic decrease with increasing mass. This can be associated with the shape of the fragments that contribute to the isomeric pair, since nuclides closer to $A$~=~132 can be assumed to have more spherical shapes and consequently, lower angular momenta. However, the angular momenta of the isotopes of Cd increase with the mass number. In order to understand this behavior, more studies are required towards the closed shell neutron configurations at $N =$~82 and the mid-shell closure at $N= $~66.

The role of single particle excitations to the angular momentum of the fission fragments can be observed in the case of In.~The deduced $J_\mathrm{rms}$ values of the odd-$Z$ In isotopes are high compared to the even-$Z$ Cd isotopes, as extra contributions to the angular momentum of the fragments are expected from the unpaired protons.
Qualitatively, this has been explained by Madsen and Brown in terms of  the polarization of the even-$Z$ core by the unpaired proton. This effect might be more prominent in the region of the $N$~=~82 spherical shell, where the deformation energy surface is more strongly governed by the protons~\cite{PhysRevLett.52.176}.~Quantitatively, Tomar~\textit{et al.}~\cite{Tomar2007} have calculated that the unpaired proton contributes $\sim$2-3 $\hbar$ to the angular momentum of the fragments in low-energy fission of actinides. Thus, a part of the angular momentum of the In isotopes can be ascribed to the odd number of protons, in agreement with observations that have been reported elsewhere~\cite{Fujiwara1982,IMANISHI1976141,Datta1986}.

\begin{figure}[h]
\centering
\includegraphics[scale=0.4]{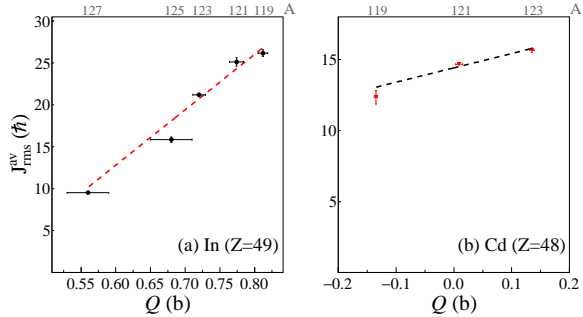}
\caption{Correlation between the electric quadrupole moments ($Q$) and the deduced $J_\mathrm{rms}$ values of the primary fission fragments for the isotopes of In in (a), and the isotopes of Cd in (b). The dashed lines represent the weighted least-square fit.}
\label{Jrms_Q}
\end{figure}

Figures.~\ref{Jrms_Q}$\color{blue}\mathrm{a}$ and~\ref{Jrms_Q}$\color{blue}\mathrm{b}$ show the correlation of the electric quadrupole moments ($Q$) of the observed products with  the angular momentum ($J^{av}_\mathrm{rms}$) of the primary fragments. The dashed lines represent the weighted least-square fit. Since our main experimental result is the fractional production rate of the high spin state, the quadrupole moments of the isotopes of Cd correspond to the values of the excited state~\cite{PhysRevLett.110.192501}, while for the isotopes of In the quadrupole moments correspond to that of the ground states~\cite{STONE20161}.~Correlations of the angular momentum of the fission fragments with the quadrupole moments of the products have also been observed by Wilhelmy~\textit{et al.}~\cite{PhysRevC.5.2041}.

The isotopes of In exhibit a strong linear correlation between  $J_\mathrm{rms}$ and $Q$, as shown in Fig.~\ref{Jrms_Q}$\color{blue}\mathrm{a}$, which suggests that the shapes of these isotopes after scission resembles the shape of the products as represented by the quadrupole moments.~Thus, either no significant changes in the shape occur during the de-excitation, or at least, after shape transitions during this process, the shapes at the initial and final stage are strongly related.~In any case, the primary fragments of In retain part of their deformation during their relaxation after scission.~Therefore, some portion of the Coulomb energy of the fragments appears as rotational kinetic energy in these fragments~\cite{PhysRev.133.B714}. Thus, additional contributions to the angular momentum of the In isotopes can arise from the electrostatic forces between the two fragments after scission, noting the effect of the Coulomb excitation in the generation of the angular momentum in the fragments~\cite{PhysRev.133.B714,RASMUSSEN1969465}.

A correlation also exists for the less deformed isotopes of Cd as can be seen in Fig.~\ref{Jrms_Q}$\color{blue}\mathrm{b}$, although not as strong. As illustrated by \mbox{Yordanov~\textit{et al.}}~\cite{PhysRevLett.116.032501}, the isomers of Cd exhibit a shape transition from oblate for $A <$~119, to almost spherical at $A =$~119~-~121, and eventually to prolate for $A>$~121, with deformation increasing with the mass number $A$.~This probably explains why the point at $A =$~119 is the one that deviates most from the least square fit. Since these nuclides are closer to a spherical configuration in shape, they are likely  to exhibit more significant shape changes during the de-excitation. 

The scission configuration for In and Cd are expected to be rather similar.~Relatively low total kinetic energy is expected for both nuclides, as the super-long fission mode is expected to dominate in the symmetric mass region.~Thus, the isotopes of In should allow for more rotational energy according to the semiclassical relation between rotational energy and angular momentum~\cite{PhysRev.93.1094}, besides the contributions due to shape deformations that were discussed earlier.~On the other hand, the isotopes of Cd should be created with higher intrinsic excitation~energy, resulting in emission of more neutrons from the primary~fragments.

In summary,  the PI-ICR technique was employed for  the first time to separate isomeric states from ground states in order to determine isomeric yield ratios by direct ion counting.~The ratios for the five isotopes of In show a decrease with respect to the mass number, while the results for the isotopes of Cd are almost independent of mass, except for the case of $A$~=~125 where a significant increase can be noticed. Moreover, by employing the code {\footnotesize TALYS} the angular momentum of the primary fragments was deduced based on the experimentally determined isomeric yield ratios.~The results for the isotopes of In reveal a rather large angular momentum, which can be attributed to the odd-$Z$ number.~A monotonic decrease of the $J_\mathrm{rms}$ values towards the proximity of the closed shell neutron configuration at $N =$ 82 can also be seen, ascribed to nuclei which are more spherical in shape and consequently carry lower angular momentum.~In addition, a linear correlation between the electric quadrupole moment of the fission products and the angular momentum of the primary fragments can be observed.~The angular momentum of the isotopes of Cd increase with mass. In order to understand this behavior more systematic studies at higher and lower masses are required.

This work was supported by the European Commission within the Seventh Framework Programme through Fission-2013-CHANDA (project No.~605203), the Swedish Radiation Safety Authority (SSM), the Swedish Nuclear Fuel and Waste Management Co. (SKB) and the Academy of Finland under the Finnish Centre of Excellence Programme 2012-2017 (Nuclear and Accelerator Based Physics Research at JYFL). A.K. acknowledges support from Academy of Finland under project No. 275389, 284516 and 312544. T.E. acknowledges support from Academy of Finland under project No. 295207. We would like to thank Prof. D. Rudolph and his group for providing us the data for $^{127}$Cd. 

\bibliographystyle{unsrt}

\end{document}